\documentclass[letterpaper, 10 pt, conference]{ieeeconf}  

    \IEEEoverridecommandlockouts                              %
    \usepackage{graphics} 
    \usepackage{epsfig} 
    \usepackage{amssymb}  
    \usepackage{graphics} 
    \usepackage{epsfig} 
    \usepackage{bm}
    \usepackage{times} 
    \usepackage{amsmath} 
    \usepackage{amssymb}  
    \usepackage{float}
    \usepackage[space, compress]{cite}
    \usepackage{mathtools}
    \usepackage[ruled,vlined,linesnumbered]{algorithm2e}
    \usepackage{graphicx}
    \usepackage{hyperref}
    \usepackage{caption}
    \usepackage{subcaption}
    \usepackage{bm,upgreek}
    \usepackage{xcolor}
    \usepackage{pbox}
    \usepackage{subcaption,graphicx}
    \usepackage{color,soul} 

    \usepackage[version=4]{mhchem}
    \usepackage{multirow}
    \usepackage{siunitx}
    
    \DeclareMathOperator*{\argmin}{arg\,min}
    
    \newcommand*\diff{\mathop{}\!\mathrm{d}}

    \title{\LARGE \bf
    Reinforcement Learning Boat Autopilot: A Sample-efficient and Model Predictive Control based Approach}
    
    \author{Yunduan Cui$^{1}$, Shigeki Osaki$^{2}$, and Takamitsu Matsubara$^{1}$
    \thanks{$^{1}$ Y. Cui and T. Matsubara are with the Division of Information Science, Graduate School of Science and Technology, Nara Institute of Science and Technology (NAIST), Japan}%
    \thanks{$^{2}$ S. Osaki is with FURUNO ELECTRIC CO., LTD., Japan}
    }

\begin{document}
    \makeatletter
    \newcommand*{\rom}[1]{\expandafter\@slowromancap\romannumeral #1@}
    \makeatother
    
    \maketitle
    \thispagestyle{empty}
    \pagestyle{empty}

    \begin{abstract}
    In this research we focus on developing a reinforcement learning system for a challenging task: autonomous control of a real-sized boat, with difficulties arising from large uncertainties in the challenging ocean environment and the extremely high cost of exploring and sampling with a real boat. To this end, we explore a novel Gaussian processes (GP) based reinforcement learning approach that combines sample-efficient model-based reinforcement learning and model predictive control (MPC). Our approach, sample-efficient probabilistic model predictive control (SPMPC), iteratively learns a Gaussian process dynamics model and uses it to efficiently update control signals within the MPC closed control loop. A system using SPMPC is built to efficiently learn an autopilot task. After investigating its performance in a simulation modeled upon real boat driving data, the proposed system successfully learns to drive a real-sized boat equipped with a single engine and sensors measuring GPS, speed, direction, and wind in an autopilot task without human demonstration.
    
    \end{abstract}
    
    \section{Introduction}\label{S1}
    
    Autonomous vehicles including smart cars and unmanned aircraft/watercraft form a rising field that not only brings beneficial changes to our transportation infrastructure such as relieving labor shortages, avoiding collisions, and providing assistance to humans \cite{fagnant2015preparing}, but also further enable ventures such as resource exploration \cite{pastore2010improving} and search and rescue \cite{tomic2012toward}. On the other hand, it is arduous to obtain good control policies for such autonomous systems since preparing human driving demonstrations encompassing a broad range of possible scenarios subject to different environmental settings is a prohibitive endeavour. Such difficulties make the use of reinforcement learning (RL) \cite{sutton1998reinforcement} an appealing prospect since RL provides a natural matter to autonomously discover optimal policies from unknown environments via trial-and-error interactions \cite{kober2013reinforcement}. {Even though RL has already been widely applied to both autonomous ground \cite{7989202} and air \cite{tran2015reinforcement} vehicles, its application to autonomous boats, or unmanned surface vehicles, remains relatively limited \cite{liu2016unmanned} primarily due to:}
    \begin{enumerate}
        \item Uncertainties in a dynamic ocean environment, e.g. frequently changing disturbances due to the unpredictable wind and current, signal noise, and hysteresis of onboard sensors, that strongly impact boat handling and dynamics.
        \item The extremely high time cost of exploring and sampling with real boats.
    \end{enumerate}
    Recent works in this area mainly focused on traditional control methods including proportional integral derivative (PID) controllers \cite{naeem2012colregs}, linear quadratic controller (LQR) \cite{lefeber2003tracking}, model predictive controllers (MPC) \cite{annamalai2015robust} and neural networks \cite{peng2013adaptive}. However these, to the best of the authors' knowledge, remain insufficient to drive a real-sized boat in an open ocean environment without human assistance and demonstration.
    
    \begin{figure}
    \begin{center}
    \includegraphics[width=1\columnwidth]{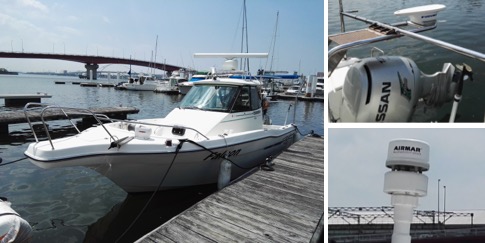}
    \caption{The Nissan Joy Fisher 25 for experiment (left) with GPS/speed/direction sensor, engine (right top), and the wind sensor (right bottom).}
    \label{fig:ship}
    \end{center}
    \end{figure}
    
    To tackle the main difficulties of autonomous boats mentioned above, an RL method should consider the large uncertainties from strong disturbance and noise while maintaining sample efficiency.
    One potential solution is the combination of the model-based RL \cite{polydoros2017survey} and Gaussian processes (GP) \cite{rasmussen2006gaussian} since model-based RL methods contribute to better data efficiency than model-free ones by learning policies from a trained model instead of directly from the environment, GP is a powerful tool that naturally takes the model's uncertainties into account. 
    {A GP based actor-critic learning model was proposed in \cite{ghavamzadeh2016bayesian}. Martin et al. applied GP to temporal difference RL to underwater robot navigation in an indoor pool \cite{Martin2018SparseGP}.}
    {
     As one state-of-the-art model-based RL method, PILCO \cite{deisenroth2013gaussian}  reduces model bias by explicitly incorporating GP model uncertainty into planning and control. Assuming the target dynamics are fully controllable, it learns an optimal policy by a long-term planning at the initial state. However, applying PILCO to an autonomous boat is difficult due to the uncontrollable and unpredictable disturbances like the wind and current. A proper feedback control against these disturbances by re-planning is computationally demanding since a large number of parameters in a state-feedback policy are optimized, while ignoring them in the long-term planning may result in an accumulated model error and bad control performances.}
    
    
    {In this research we develop a novel RL approach specialized for autonomous boat control, specifically, automatically driving a real boat in an autopilot task. Our approach, sample-efficient probabilistic model predictive control (SPMPC), iteratively learns a Gaussian process dynamics model to increase the robustness of control against several unpredictable and frequently changing noises and disturbances and efficiently optimizes control signals under a close-loop MPC to reduce the heavy computation cost of full-horizon planning in \cite{deisenroth2013gaussian}. A system based on SPMPC is built to learn autopilot task with a real-size boat equipped with a single engine and sensors of GPS, speed, direction, and wind (Fig. \ref{fig:ship}). The results show the capability of the proposed approach in autonomous boats with both robustness to disturbances and sample efficiency. Compared with prior works, the contributions of this paper are the following:
    \begin{enumerate}
        \item {Propose a sample-efficient, probabilistic model-based RL suitable for real boat control.}
        \item Build a learning control system with bias compensation for the boat autopilot task. Its capabilities are evaluated in a simulation tuned with real boat driving data.
        \item Conduct experimental validation of the system on a real-size boat in a real ocean environment.
    \end{enumerate}
    }
    
    

    The remainder of this paper is organized as follows. Section \ref{S2} presents related works. Section \ref{S3} introduces SPMPC. Section \ref{S4} details the RL system for autonomous boats. The results of simulation and real autopilot tasks are in Section \ref{S5}. Discussions follow in Section \ref{S6}.
    
    {
    \section{Related Works}\label{S2}

    Williams et al. \cite{7989202} combined MPC with model-based RL and successfully implement it to autonomous ground vehicles. {This work approximates the dynamics model by means of neural networks that are difficult to follow the fully Bayesian formalism. Thus, it would require a large number of samples for model learning and hyper-parameter selection.} Cao et al. \cite{cao2017gaussian} proposed a GP model-based MPC controller to support state uncertainties in an unmanned quadrotor simulation control problem, utilizing a robust MPC controller which requires pre-collected training data to learn the GP model without self-exploration for data collection. Kamthe et al. \cite{kamthe2018data} first extend PILCO to avoid the full-horizon planning in model-based RL by introducing MPC to moderate the real-time disturbances within a closed control loop. This work successfully shows its sample efficiency in cart-pole and double pendulum in simulation tasks without considering external disturbances. {One potential limitation of \cite{kamthe2018data} towards real-world challenging control problems may be the relatively heavy computational cost since its optimization is executed with a dimensionality expanded deterministic dynamical system with Lagrange parameter and state constraint under the Pontryagin's maximum principle (PMP).} {In this work, we focus on the RL system for real-world autonomous boats where state-constraints are less important in the currently focused problems, rather a simpler and more computationally efficient method would be desirable.}
    Our proposed approach, SPMPC, has a certain similarity to the method in \cite{kamthe2018data} with the following features as advantages for real boat autopilot:
    \begin{enumerate}
        \item By separating the uncertain state and deterministic control signal in prediction (Section \ref{S3-2}), our approach directly optimizes the long-term cost with neither the expanded dynamics nor state constraints, therefore is more computationally efficient and suitable for real-world control problems. 
        \item Our approach employs {\it bias compensation} to compensate the bias, error between the initial states in the control plan and the actual ship before control execution due to the relatively-low control frequency. 
    \end{enumerate}
    }

    \section{Approach}\label{S3}
    {In this section, we detail SPMPC. As a model-based RL, it stores the knowledge of the environment in a learned model (Section \ref{S3-1}). Its control policy is made by combining the long-term optimization (Section \ref{S3-2}) and the MPC controller (Section \ref{S3-3}). Its RL loop follows:
    \begin{enumerate}
        \item Observe state and optimize control signals based on the model and policy (long-term optimization $+$ MPC)
        \item Operate the control signal to the system, observe the next state and collect the sample
        \item Update the model with all samples, return to 1)
    \end{enumerate}
    }
    
    \subsection{Model Learning}\label{S3-1}
    {In this work, a dynamical system is modeled by GP:}
    \begin{gather}    
    \bm{x}_{t+1} = f(\bm{x}_{t}, \bm{u}_{t}) + \bm{w}, \quad \bm{w} \sim \mathcal{N}(\bm{0}, \bm{\Sigma}_{w}),
    \label{eq_dynamic_system}
    \end{gather}
    {$\bm{x}\in \mathbb{R}^{D}$ is state, $\bm{u}\in \mathbb{R}^{U}$ is control signal, $f$ is the unknown transition function, $\bm{w}$ is an i.i.d. system noise. Note that in this work the dimension of output state $\bm{x}_{t+1}$ can be smaller than input $\bm{x}_t$ to handle uncontrollable states, e.g., wind, we assume they all have dimension $D$ for the simplification of equations. Given the training input tuples $\tilde{\bm{x}}_{t}:=(\bm{x}_{t}, \bm{u}_{t})$, and their training targets $\bm{y}_t := \bm{x}_{t+1}$, for each target dimension $a = 1, ..., D$, a GP model is trained based on the latent function $y^{a}_{t} = f_{a}(\tilde{\bm{x}}_{t}) + w_{a}$ with a mean function $m_{f_{a}}(\cdot)$ and squared exponential covariance kernel function:}
    \begin{gather}   
    k_a(\tilde{\bm{x}}_{i}, \tilde{\bm{x}}_{j})=\alpha^{2}_{f_{a}}\!\exp\!\big(\!\!-\frac{1}{2}(\tilde{\bm{x}}_{i} - \tilde{\bm{x}}_{j})^{T}\tilde{\bm{\varLambda}}_{a}^{-1}(\tilde{\bm{x}}_{i} - \tilde{\bm{x}}_{j})\big),
    \label{eq_kernel}
    \end{gather}
    {where $\alpha^{2}_{f_{a}}$ is the variance of $f_{a}$, and $\tilde{\bm{\varLambda}}_{a}$ is the diagonal matrix of training inputs' length scales in kernel. The parameters of GP model are learned by evidence maximization \cite{rasmussen2006gaussian,mackay2003information}. Define $\tilde{\bm{X}} = [\tilde{\bm{x}}_{1}, ..., \tilde{\bm{x}}_{N}]$ as the training inputs set, ${\bm{Y}}^{a} = [\bm{y}^{a}_{1}, ..., \bm{y}^{a}_{N}]$ as the collection of the training targets in corresponding dimension, $\bm{k}_{a, *}=k_{a}(\tilde{\bm{X}}, \tilde{\bm{x}}_{*})$, $k_{a, **}=k_{a}(\tilde{\bm{x}}_{*}, \tilde{\bm{x}}_{*})$, $K_{i,j}^{a}=k_{a}(\tilde{\bm{x}}_{i}, \tilde{\bm{x}}_{j})$ as the corresponding element in $\bm{K}^{a}$, and $\bm{\beta}_{a}=(\bm{K}^{a}+\alpha^{2}_{f_{a}}\bm{I})^{-1}\bm{Y}^{a}$, the GP predictive distribution of a new input $\tilde{\bm{x}}_{*}$ follows:} 
    \begin{gather}   
    p(f_{a}(\tilde{\bm{x}}_{*})| \tilde{\bm{X}}, {\bm{Y}}^{a})=\mathcal{N}(f_{a}(\tilde{\bm{x}}_{*})|m_{f_{a}}(\tilde{\bm{x}}_{*}), \sigma^{2}_{f_{a}}(\tilde{\bm{x}}_{*})),\label{eq_GP_m_V_1}\\
    m_{f_{a}}(\tilde{\bm{x}}_{*}) = \bm{k}_{a, *}^{T}(\bm{K}^{a}+\alpha^{2}_{f_{a}}\bm{I})^{-1}\bm{Y}^{a} = \bm{k}_{a, *}^{T}\bm{\beta}_{a},\label{eq_GP_m_V_2}\\
    \sigma^{2}_{f_{a}}(\tilde{\bm{x}}_{*}) = k_{a, **} - \bm{k}_{a, *}^{T}(\bm{K}^{a}+\alpha^{2}_{f_{a}}\bm{I})^{-1}\bm{k}_{a, *}.
    \label{eq_GP_m_V_3}
    \end{gather}
    
    
    
    \subsection{Optimization of an Open-loop Control Sequence}\label{S3-2}
    
    Define a one-step cost function $l(\cdot)$, the next step is to search a multiple steps optimal open-loop control sequence $\bm{u}^{*}_{t}, ..., \bm{u}^{*}_{t+H-1}$ that minimizes the expected long-term cost:
    \begin{gather}  
    \begin{split}
    [\bm{u}^{*}_{t}, ..., \bm{u}^{*}_{t+H-1}] = \argmin_{\bm{u}_{t}, ..., \bm{u}_{t+H-1}}\sum^{t+H-1}_{s = t} \gamma^{s-t} l(\bm{x}_{s}, \bm{u}_{s}),
    \label{eq:find_control}
    \end{split}
    \end{gather}
    where $\gamma \in [0, 1]$ is the discount factor. 
    {
    In general, including the model uncertainties in Eq. \ref{eq:find_control} is difficult since approximating the intractable marginalization of model input by traditional methods like Monte-Carlo sampling is computationally demanding, especially for each candidate of control sequence during optimization.
    As one solution, we utilize analytic movement matching \cite{girard2003gaussian,deisenroth2009analytic} that assumes a Gaussian model input and provides exact analytical expressions for the mean and variance. In this section, we propose a modified moment-matching to efficiently optimize the deterministic control sequence in Eq. \ref{eq:find_control} by separating the uncertain state and deterministic control in the prediction:
    }
    \begin{gather}  
    \begin{split}
    [\bm{\mu}_{t+1}, \bm{\Sigma}_{t+1}] = f(\bm{\mu}_{t}, \bm{\Sigma}_{t}, \bm{u}_{t}).
    \label{MM}
    \end{split}
    \end{gather}
    Based on the moment-matching \cite{deisenroth2009analytic}, the proposed method starts by separating the kernel function in Eq. (\ref{eq_kernel}) by assuming the state and control signal are independent:
    \begin{align}
    \begin{split}    
    k_a(\bm{x}_{i}, \bm{u}_{i}, \bm{x}_{j}, \bm{u}_{j}) = k_a(\bm{u}_{i}, \bm{u}_{j})\times k_a(\bm{x}_{i}, \bm{x}_{j}).
    \label{eq_kernel_mpc}
    \end{split}
    \end{align}
    Define $\bm{k}_{a}(\bm{u}_{*}) = k_{a}(\bm{U}, \bm{u}_{*})$ and $\bm{k}_{a}(\bm{x}_{*}) = k_{a}(\bm{X}, \bm{x}_{*})$, the mean and covariance related to Eqs. (\ref{eq_GP_m_V_2}) and (\ref{eq_GP_m_V_3}) then follows:
    \begin{align}
    \begin{split}    
    m_{f_{a}}(\bm{x}_{*}, \bm{u}_{*}) = \big(\bm{k}_{a}(\bm{u}_{*})\times\bm{k}_{a}(\bm{x}_{*})\big)^{T}\bm{\beta}_{a},
    \label{eq_GP_m_mpc}
    \end{split}
    \end{align}
    \begin{align}
    \begin{split}    
    &\sigma^{2}_{f_{a}}(\bm{x}_{*}, \bm{u}_{*}) = \big(k_{a}(\bm{u}_{*}, \bm{u}_{*})\times k_{a}(\bm{x}_{*}, \bm{x}_{*})\big) - \\&(\bm{k}_{a}(\bm{u}_{*})\!\!\times\!\bm{k}_{a}(\bm{x}_{*}))^{T}(\bm{K}^{a}\!\!+\!\alpha^{2}_{f_{a}}\bm{I})^{-1}(\bm{k}_{a}(\bm{x}_{*})\!\!\times\!\bm{k}_{a}(\bm{u}_{*}))
    \label{eq_GP_v_mpc}
    \end{split}
    \end{align}
    where the element in $\bm{K}^{a}$ is $K_{i,j}^{a}=k_{a}(\bm{x}_{i}, \bm{x}_{j})\times k_{a}(\bm{u}_{i}, \bm{u}_{j})$ and $\bm{\beta}_{a}=(\bm{K}^{a}+\alpha^{2}_{f_{a}}\bm{I})^{-1}\bm{Y}^{a}$, $\bm{X} = [\bm{x}_{1}, ..., \bm{x}_{N}]$ and $\bm{U} = [\bm{u}_{1}, ..., \bm{u}_{N}]$ are the training inputs, ${\bm{Y}}^{a} = [\bm{y}^{a}_{1}, ..., \bm{y}^{a}_{N}]$ is the training targets in corresponding dimension.
    Given the uncertain state $\bm{x}_{*}\sim \mathcal{N}(\bm{\mu}, \bm{\Sigma})$ and deterministic control signal $\bm{u}_{*}$ as inputs, the predicted mean follows: 
    \begin{gather}
    \begin{split}    
    p(f(\bm{x}_{*}, \bm{u}_{*})|\bm{\mu}, \bm{\Sigma}, \bm{u}_{*}) \approx\mathcal{N}(\bm{\mu}_{*}, \bm{\Sigma}_{*}),
    \label{eq_mm_mpc}
    \end{split}
    \end{gather}
    \begin{gather}
    \begin{split}    
    \mu_{a*}&= \bm{\beta}_{a}^{T}\bm{k}_{a}(\bm{u}_{*})\!\!\int\!\! \bm{k}_{a}(\bm{x}_{*}) p(\bm{x}_{*}|\bm{\mu}, \bm{\Sigma}) \diff \bm{x}_{*} = \bm{\beta}_{a}^{T}\bm{l}_{a}.
    \label{eq_mm_m_mpc}
    \end{split}
    \end{gather}
    For target dimensions $a, b = 1, ..., D$, the predicted variance $\Sigma_{aa*}$ and covariance $\Sigma_{ab*}, a \ne b$ follow:
    \begin{align}
    \begin{split}    
    \Sigma_{aa*}&=\bm{\beta}_{a}^{T}\bm{L}\bm{\beta}_{a} \!+\! \alpha^{2}_{f_{a}} \!-\! tr\big((\bm{K}^{a} \!+\! \sigma_{w_{a}}^2\bm{I})^{-1}\bm{L}\big) \!-\! \mu_{a*}^{2},
    \label{eq_mm_var_mpc}
    \end{split}
    \end{align}
    \begin{align}
    \begin{split}    
    \Sigma_{ab*} &=\bm{\beta}_{a}^{T}\bm{Q}\bm{\beta}_{b} - \mu_{a*}\mu_{b*}.
    \label{eq_mm_covar_mpc}
    \end{split}
    \end{align}
    Vector $\bm{l}_{a}$ and matrices $\bm{L}$, $\bm{Q}$ has elements:
    \begin{align}
    \begin{split}    
    l_{ai} &= k_{a}(\bm{u}_{i}, \bm{u}_{*})\int k_{a}(\bm{x}_{i}, \bm{x}_{*}) p(\bm{x}_{*}|\bm{\mu}, \bm{\Sigma}) \diff \bm{x}_{*}\\
    &= k_{a}(\bm{u}_{i}, \bm{u}_{*})\alpha^{2}_{f_{a}}|\bm{\Sigma}\bm{\varLambda}_{a} + \bm{I}|^{-\frac{1}{2}}\\ &\times\exp\big(-\frac{1}{2}(\bm{x}_{i}-\bm{\mu})^T(\bm{\Sigma}+\bm{\varLambda}_{a})^{-1}(\bm{x}_{i}-\bm{\mu})\big).
    \label{eq_mm_l_mpc}
    \end{split}
    \end{align}
    \begin{align}
    \begin{split}
    L_{ij} 
    &= k_{a}(\bm{u}_{i},\bm{u}_{*})k_{a}(\bm{u}_{j},\bm{u}_{*})\frac{k_{a}(\bm{x}_{i},\bm{\mu})k_{a}(\bm{x}_{j},\bm{\mu})}{|2\bm{\Sigma}\bm{\varLambda}_{a}^{-1} + \bm{I}|^{\frac{1}{2}}}\\
    &\times \!\exp\!\big((\bm{z}_{ij}\!-\!\bm{\mu})^{T}(\bm{\Sigma}\!+\!\frac{1}{2}\bm{\varLambda}_{a})^{-1}\bm{\Sigma}\bm{\varLambda}_{a}^{-1}(\bm{z}_{ij}\!-\!\bm{\mu})\big),
    \label{eq_mm_L_mpc}
    \end{split}
    \end{align}
    \begin{align}
    \begin{split}
    Q_{ij} &= \alpha_{f_{a}}^{2}\alpha_{f_{b}}^{2}k_{a}(\bm{u}_{i},\bm{u}_{j})k_{b}(\bm{u}_{i},\bm{u}_{j})|(\bm{\varLambda}_{a}^{-1}\!+\!\bm{\varLambda}_{b}^{-1})\bm{\Sigma}\!+\!\bm{I}|^{-\frac{1}{2}}\\
    &\times\!\exp\!\big(\!-\!\frac{1}{2}(\bm{x}_{i} \!-\! \bm{x}_{j})^{T}(\bm{\varLambda}_{a} \!+\! \bm{\varLambda}_{b})^{-1}(\bm{x}_{i} \!-\! \bm{x}_{j})\big)\\
    &\times\!\exp\!\big(\!-\!\frac{1}{2}(\bm{z}_{ij}^{\prime}\!-\!\bm{\mu})^{T}\bm{R}^{-1}(\bm{z}_{ij}^{\prime}\!-\!\bm{\mu})\big)
    \label{eq_mm_Q_mpc}
    \end{split}
    \end{align}
    where $\bm{\varLambda}_{a}$ is the diagonal matrix of training inputs’ length scales in kernel $k_a(\bm{x}_{i}, \bm{x}_{j})$. $\bm{z}^{\prime}$ and $\bm{R}$ are given by:
    \begin{align}
    \begin{split}
    \bm{z}^{\prime}_{ij} = \tilde{\bm{\varLambda}}_{b}(\bm{\varLambda}_{a}+\bm{\varLambda}_{b})^{-1}\bm{x}_{i} + \bm{\varLambda}_{a}(\bm{\varLambda}_{a}+\bm{\varLambda}_{b})^{-1}\bm{x}_{j},
    \label{eq_mm_z}
    \end{split}
    \end{align}
    \begin{align}
    \begin{split}
    \bm{R}=(\bm{\varLambda}_{a}^{-1}+\bm{\varLambda}_{b}^{-1})^{-1}+\bm{\Sigma}_{t}.
    \label{eq_mm_R}
    \end{split}
    \end{align}
    
    {Compared with \cite{kamthe2018data} that drives a dimensionality expanded system in the moment-matching, our approach focused on simplifying the moment-matching with uncertain state and deterministic control signal towards an efficient optimization. Given the cost function $l(\cdot)$, prediction length $H$, initial control sequence $\bm{u}^{0}_{t}, ..., \bm{u}^{0}_{t+H-1}$, and control constraints, any constrained nonlinear optimization method can search for the optimal control sequence to minimize the long-term cost in Eq. \ref{eq:find_control} where the future states and corresponding uncertainties in $H-1$ steps are predicted by the modified moment-matching. In this work, sequential quadratic programming (SQP) \cite{nocedal2006sequential} is used.}
    

    \subsection{Model Predictive Control Framework}\label{S3-3}
    
    
    \begin{figure}
    \begin{center}
    \includegraphics[width=1\columnwidth]{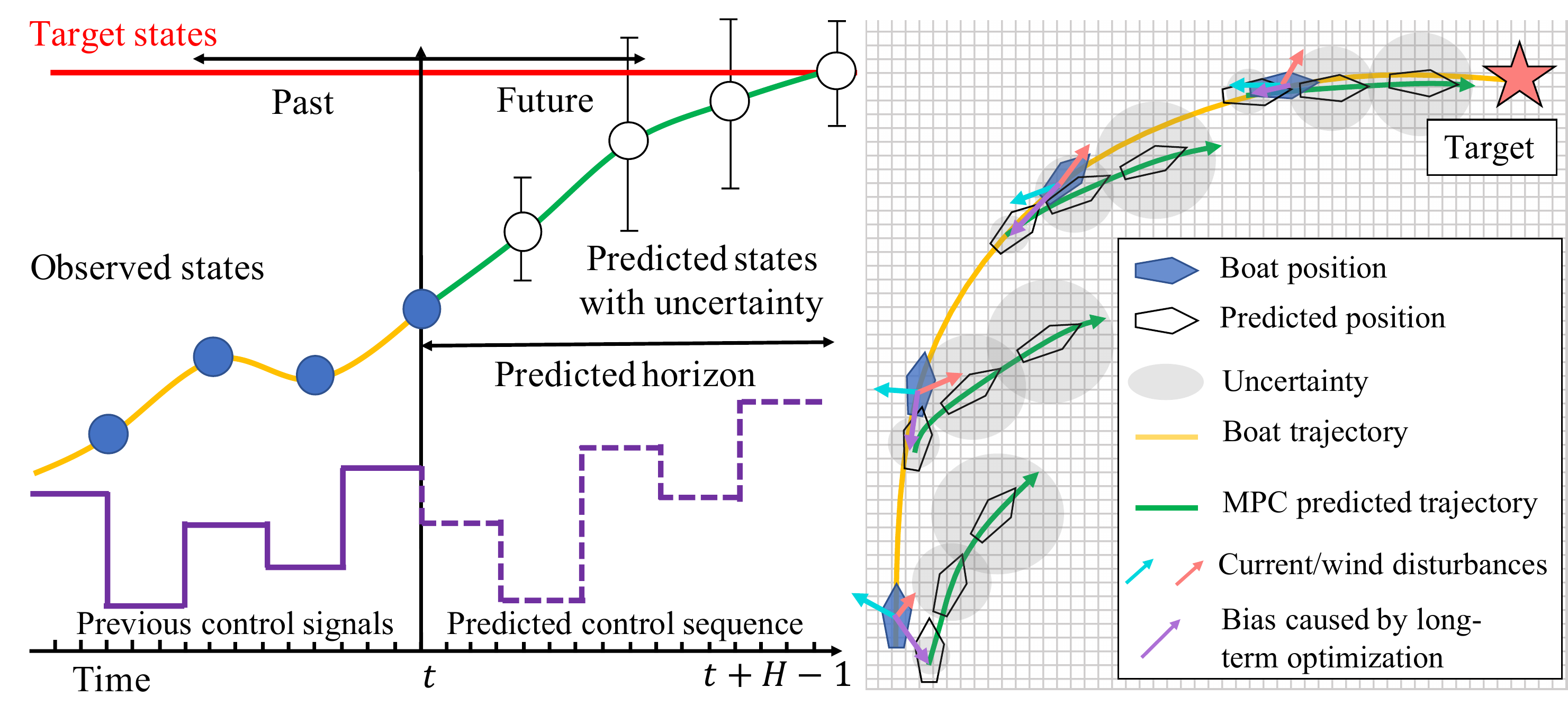}
    \caption{Overall of the MPC framework. Left: The MPC framework. Right: the MPC framework with long-term prediction in autonomous boat control.}
    \label{fig:mpc_ship}
    \end{center}
    \end{figure}
    
    
    After obtaining the optimal $H$-step open-loop control trajectory $\bm{u}^{*}_{t}, ..., \bm{u}^{*}_{t+H-1}$ by minimizing the long-term cost in Eq. \ref{eq:find_control} (Section \ref{S3-2}), the next step is to transfer this open-loop control sequence to an implicit feedback controller using the MPC framework \cite{mayne2000constrained}. As shown in the left side of Fig. \ref{fig:mpc_ship} for each step $t$, given the current state $\bm{x}_{t}$, an optimal $H$-step open-loop control sequence is determined following Eq. \ref{eq:find_control} to minimize the long-term cost based on a $H$-step prediction using Eq. \ref{MM}. The first control signal $\bm{u}^{*}_{t}$ is then applied to the system and then get the next step state $\bm{x}_{t+1}$. An implicit closed-loop controller is then obtained by re-planning the $H$-step open-loop control sequence at each coming state.

    \begin{table}
    \caption{The observed states and control parameters of our autonomous boat system}
    \begin{center}
    \subcaption*{Observed State}
    \begin{tabular}{|p{1cm}|p{4cm}|p{2cm}|}
    \hline
    Name & Description & From \\
    \hline
    X & The position in X axis & GPS sensor \\
    \hline
    Y & The position in Y axis & GPS sensor \\
    \hline
    ss & Boat speed & Direction sensor \\
    \hline
    sd & Boat direction & Direction sensor\\
    \hline
    rws & Relative wind speed  & wind sensor \\
    \hline
    rwd & Relative wind direction & wind sensor \\
    \hline
    \end{tabular}
    \bigskip
    \subcaption*{Control Signal}
    \begin{tabular}{|p{1cm}|p{4cm}|p{2cm}|}
    \hline
    Name & Description & Range \\
    \hline
    RR & The steering angle & $[-30, 30]^{\circ}$ \\
    \hline
    Throttle & The throttle value of engine & $[-8000, 8000]$\\
    \hline
    \end{tabular}
    \label{tab:sim_state_action}
    \end{center}
    \end{table}

    \section{Autonomous Boat Control System }\label{S4}
    In this section, the details of building a system specialized for real boat autopilot using the SPMPC are introduced.
    
    \subsection{Autonomous Boat System}\label{S4-1}
    As shown in Fig. \ref{fig:ship}, the proposed system was applied to a real Nissan JoyFisher 25 (length: 7.93 m, width: 2.63 m, height: 2.54 m) fitted with a single Honda BF130 engine and two sensors: a Furuno SC-30 GPS/speed/direction sensor and a Furuno WS200 wind sensor. As noted in Table \ref{tab:sim_state_action}, observed states include the boat's current GPS position, speed, and direction from the direction sensor, and relative wind speed and direction from the wind sensor. {The boat was not equipped with a current sensor.} Control parameters are defined as the steering angle and engine throttle that respectively control the boat's turning and linear velocity. {There are two disturbances that strongly affect the navigation of our system in the real ocean: the unobservable ocean current and the observable but unpredictable wind. In this work, we build a SPMPC system based on our proposed approach to alleviate the effect of these disturbances.}
    
    \begin{figure}
    \begin{center}
    \includegraphics[width=1\columnwidth]{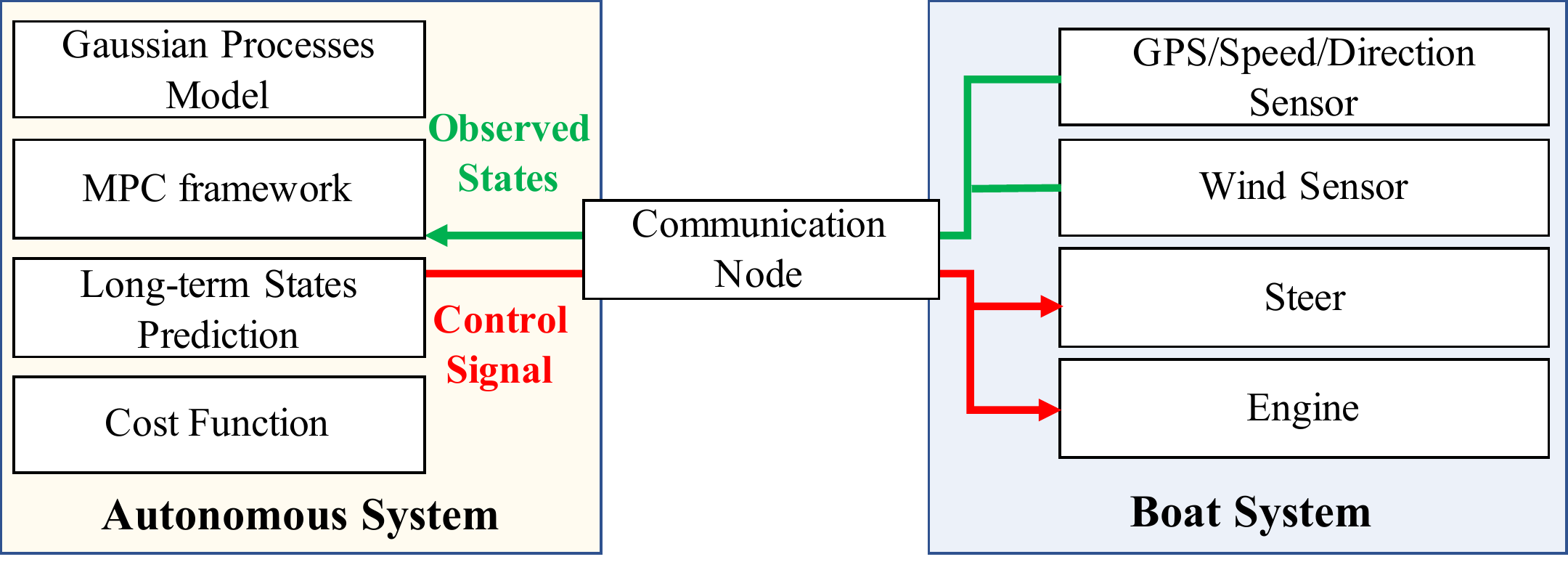}
    \caption{The SPMPC System.}
    \label{fig:real_system}
    \end{center}
    \end{figure}

    \subsection{SPMPC System}\label{S4-2}
    
    The overall autonomous system that applies SPMPC to autonomous boat control has three parts: training the GP model, predicting long-term states, and running the MPC framework. The GP model is trained following Section \ref{S3-1}. {Since the wind is an unpredictable and uncontrollable disturbance that strongly affects the boat behaviors, the dimensions of input and output in the GP model are different. We define the input states as $\bm{x}_{t} = [X_{t}, Y_{t}, ss_{t}, sd_{t}, rws_{t}\times\sin(rwd_{t}), rws_{t}\times\cos(rwd_{t})]$ where the relative wind speed and direction are translated to a 2D vector, the training targets as $\bm{y}_{t}=[X_{t+1}, Y_{t+1}, ss_{t+1}, sd_{t+1}]$. The control signal is defined as $\bm{u}_{t} = [RR_{t}, throttle_{t}]$. Additionally, the wind states in $\bm{x}_{s}$ are fixed to their initial observation in $\bm{x}_{t}$ when optimizing the control sequence in Eq. \ref{eq:find_control}, i.e. we assume the wind is fixed during the long-term prediction.}

    {The effects of disturbances from current and the assumption of fixed wind are alleviated by the MPC framework (Section \ref{S3-3}) in the SPMPC system. As shown in the right side of Fig. \ref{fig:mpc_ship}, to avoid accumulated prediction error, the SPMPC controller updates the state affected by disturbances and re-plans a control sequence at every step. Therefore, it iteratively makes the boat reaching the target even though the predicted trajectories (green arrow) are biased compared with the real one (yellow arrow) at each step.}
    
    The complete workflow is shown in Fig. \ref{fig:real_system}. There is a software node to handle communications between the autonomous system and lower-level boat hardware interface. At each step $t$, this node first passes sensor readings to the autonomous system as the initial state following Section \ref{S4-2}. The autonomous system searches a $H$-step control signal sequence that minimizes the long-term cost based on a $H$-step state prediction subject to constant wind states. The first control signal $\bm{u}^{*}_{t}$ in the output sequence is sent to the hardware interface to control the steering and engine throttle. 
    
    \subsection{Bias Compensation for MPC}\label{S4-3}
    
    {In real-world applications, under the effect of the previous control signal $\bm{u}_{t-1}$ which is continuously sent to the system during the process of Eq. \ref{eq:find_control}, the state after obtaining $\bm{u}^{*}_{t}$ , defined as $\bm{x}^{*}_{t}$, will be different to the state used for searching $\bm{u}_{t}$, defined as $\bm{x}^{\prime}_{t}$. The bias $\bm{b}_{t}= \bm{x}^{*}_{t} - \bm{x}^{\prime}_{t}$, shown as the purple arrows in the right side of Fig. \ref{fig:mpc_ship}, will worsen the controller's performance, especially when optimization is lengthy. Therefore, a bias compensation is required to mitigate this effect by predicting the boat's state after each optimization.
    {Denoting the optimization time with $t_{opt}$, we implement the bias compensation to predict position $[X_{bias}, Y_{bias}]$ after $t_{opt}$ according to the boat's current position $[X_{}, Y_{}]$, speed $ss$ and direction $sd$: 
    \begin{gather}  
    \begin{split}
    X_{bias} = X + ss \times \sin{(sd)} \times t_{opt} \\
    Y_{bias} = Y + ss \times \cos{(sd)} \times t_{opt}
    \label{bias}
    \end{split}
    \end{gather}
    It will be used as the initial state in Eq. \ref{eq:find_control} to reduce the effect of the control signals during optimization.
    }
    
    \subsection{Learning Process}\label{S4-4}
    
    {Here we introduce the learning process of SPMPC system. Following Table \ref{tab:sim_state_action}, the GP model is initialized by $N_{trials}\times L_{rollout}$ samples with random actions as shown in lines 1-10 in Algorithm \ref{alg:1}. At step $j$, the state $\bm{x}_j$ is observed via the function ReadSensor(), and the random control signal $\bm{u}_j$ is generated with RandAct(). After applying the control signal to the system, the target $\bm{y}_j$ is observed. $\bm{x}_j$ and $\bm{y}_j$ are added to the GP model's training input/target sets.} 
    
    {The next is the RL process (line 11 to 22), which runs $N_{trials}$ rollouts to iteratively improve performance. At each step, the current state $\bm{x}^{\prime}$ is first observed, and the state with bias compensation, $\bm{x}^{*}=[X_{bias}, Y_{bias}]$ following Eq. \ref{bias} is predicted by BiasComp(). The control signal $\bm{u}_{j}$ is calculated by Eq. \ref{eq:find_control} via function OptActions() where $H$ is the MPC prediction step, $costFun$ is the cost function for the activity being undertaken. The corresponding state and target are then observed and added to the training input/target sets. The GP model is updated after each rollout.
    }

    \begin{algorithm}[t]
        \SetKwFunction{Reset}{ResetBoat}
        \SetKwFunction{GRA}{RandAct}
        \SetKwFunction{OPA}{OptAct}
        \SetKwFunction{RS}{ReadSensor}
        \SetKwFunction{OA}{OperateActions}
        \SetKwFunction{TGP}{TrainGP}
        \SetKwFunction{BC}{BiasComp}
    
        \SetKwFunction{UpPN}{UpdatePNetwork}
        \SetKwProg{Fn}{$Function$}{:}{\KwRet}
        \KwIn{$N_{initial}, N_{trials}, L_{rollout}, H, costFun$}  
        \KwOut{$model$}
        \SetKwFunction{FMain}{SPMPC}
        \SetKwProg{Fn}{Function}{:}{}
        \Fn{\FMain{$N_{initial}, ..., costFun$}}
        {
        \For{$ i = 1, 2, ..., N_{initial}$}
        {
            \Reset{}\\
            \For{$ j = 1, 2, ..., L_{rollout}$}
            {   
                $\bm{x}_{j}$ = \RS{}\\
                $\bm{u}_{j}$ = \GRA{}\\
                \OA{$\bm{u}_{j}$}\\
                $\bm{y}_{j}$ = \RS{}\\
                $\tilde{\bm{x}}_{j} = (\bm{x}_{j}, \bm{u}_{j})$\\ 
                $\tilde{\bm{X}}=\{\tilde{\bm{X}}, \tilde{\bm{x}}_{j}\}, \bm{Y}=\{\bm{Y}, \bm{y}_{j}\}$
            }
        }
        $model$ = \TGP{$\tilde{X}, Y$}\\
        \smallskip
        \For{$ i = 1, 2, ..., N_{trials}$}
        {
            \Reset{}\\
            \For{$ j = 1, 2, ..., L_{rollout}$}
            {   
                $\bm{x}^{\prime}$ = \RS{}\\
                $\bm{x}^{*}$ = \BC{$\bm{x}^{\prime}$}\\
                $\bm{u}_{j}$ = \OPA{$\bm{x}^{*}, H, model, costFun$}\\
                $\bm{x}_{j}$ = \RS{}\\
                \OA{$\bm{u}_{j}$}\\
                $[\bm{y}_{j}]$ = \RS{}\\
                $\tilde{\bm{x}}_{j} = (\bm{x}_{j}, \bm{u}_{j})$\\            
                $\tilde{\bm{X}}=\{\tilde{\bm{X}}, \tilde{\bm{x}}_{j}\}, \bm{Y}=\{\bm{Y}, \bm{y}_{j}\}$
            }
            $model$ = \TGP{$\tilde{X}, Y$}\\
        }
        \textbf{return} $model$ 
        }
        \textbf{End Function}
      \caption{{SPMPC} for boat autopilot}
      \label{alg:1}
      \end{algorithm}

    \section{Experiments}\label{S5}
    \subsection{Simulation Experiments}\label{S5-1}
    {The SPMPC system is first investigated in a simulator developed by FURUNO ELECTRIC CO., LTD following the state definition and the control constraints in Table \ref{tab:sim_state_action}. It simulates feasible boat behaviors based on real driving data of the Nissan Joy Fisher 25 in the real ocean environment.} In the simulation, we set an autopilot task in a $500 \times 500$ $m^{2}$ open area with arbitrary wind and ocean current. {The objective is to drive the boat from its initial position $[0, 0]$ to the target position $P_{target} = [400, 250]$ and remain as close as possible.} {The states and actions are defined in Table \ref{tab:sim_state_action}.} In terms of environmental parameters, the wind and current directions follow a uniform distribution $[wd, cd] \sim U(-180, 180)^{\circ}$, wind speed $ws \sim U(0, 10)$ $m/s$, and current speed $cs \sim U(0, 1)$ $m/s$. At each step both wind and current slightly change to simulate an open oceanic environment following $[\Delta wd, \Delta cd] \sim U(-0.1, 0.1)^{\circ}$, $[\Delta ws, \Delta cs] \sim U(-0.1, 0.1)$ $m/s$. The control signal is operated over the course of $3.5$ s, consisting of a $2.5$ s operation time and $1$ s optimization time $t_{opt}$ for bias compensation. Optimization stops when reaching its termination criteria, with the time taken recorded as an evaluation criterion.
    
    \subsubsection{Cost Function Definition}\label{S5-1-1}
    
    Two one-step cost functions in Eq. \ref{eq:find_control} are implemented to consider the effect of the predicted state's uncertain information in cost function. One is based on the Euclidean-distance between the predicted mean of boat position in the $s$-th step $\bm{P}_{s} = [X_{s}, Y_{s}]$ and $\bm{P}_{target}$:
    \begin{gather}  
    \begin{split}
    l(\bm{P}_s) =  \frac{1}{2}||\bm{P}_s-\bm{P}_{target}||^{2}.
    \label{deterministic_cost_function}
    \end{split}
    \end{gather}
    Another one is based on Mahalanobis-distance that considers both predicted mean and variance of boat position:
    \begin{gather}  
    \begin{split}
    l(\bm{P}_s, \bm{\Sigma}_{\bm{P}_{s}}) \!=\! \frac{1}{2}(\bm{P}_s\!-\!\bm{P}_{target})^{T}\tilde{\bm{S}}(\bm{P}_s\!-\!\bm{P}_{target})
    \label{probabilistic_cost_function}
    \end{split}
    \end{gather}
    where $\tilde{\bm{S}}=\bm{W}^{-1}(\bm{I}+\bm{\Sigma}_{\bm{P}_{s}}\bm{W}^{-1})^{-1}$, $\bm{W}$ is a diagonal matrix scaled by parameter $\frac{1}{\sigma_{c}^2}$. We set $\sigma_{c} = 1$ in this experiment. Since the predicted variance in the previous step is required in Eqs. \ref{eq_mm_m_mpc} and \ref{eq_mm_var_mpc}, both two cost functions can benefit from the uncertainty information in long-term prediction.

    \begin{figure}
    \begin{center}
    \includegraphics[width=1.0\columnwidth]{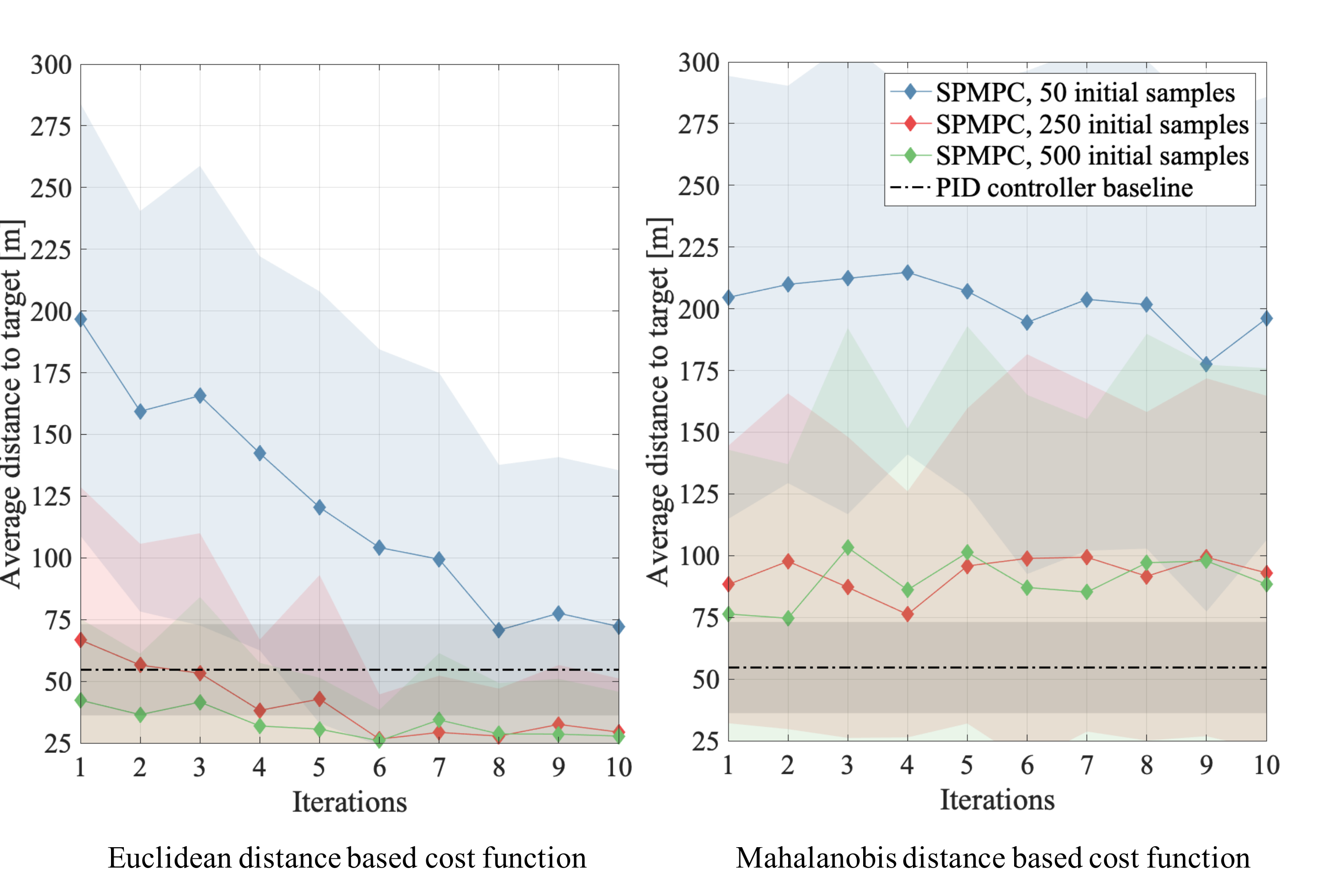}
    \caption{Convergence of SPMPC with Euclidean-distance and Mahalanobis-distance based cost functions.}
    \label{fig:cov_dist}
    \end{center}
    \end{figure}
    
    \subsubsection{Evaluation of Learning Performances}\label{S5-1-2}

    The first experiment is to investigate the convergence of SPMPC with the different type of cost function and the different number of samples available. Following Algorithm \ref{alg:1}, we set the rollout length $L_{rollout}=50$ and RL iterations $N_{trials}=10$. The MPC prediction horizon $H=5$, $N_{initial}=[1, 5, 10]$, i.e. build the initial GP model with $50$, $250$, and $500$ random samples, are tested. {The random samples for initializing GP model are generated by randomly selecting control signal $RR \sim U(-30, 30)^{\circ}$ and $throttle\sim U(-8000, 8000)$ at each step, under the random wind and current generated following Section \ref{S5-1}}. Sparse GP \cite{snelson2006sparse} with $50$ pseudo-inputs are utilized for efficient calculation. After each iteration, the GP model is tested in $10$ independent rollouts. {The discount factor is not such a sensitive parameter since the prediction horizon in SPMPC is set to be relatively short due to the large disturbances. In this experiment, we set $\gamma = 0.95$.} The learning performance was averaged over 10 times experiments.
    A comparative baseline is conducted by applying two PID controllers for steering angle and throttle value of engine. The PID error is defined as the angle between boat head and the target, and the distance from boat to the target. {The control frequency of PID controllers is set to $20$ Hz, while SPMPC operates each action over $3.5$ s, about $0.28$ Hz}. Both two PID controllers' parameters were manually tuned to $P = 1.0 , I = 0.1, D = 0.1$ based on its average performance on $1000$ trials with the same environmental parameters above.

    Figure \ref{fig:cov_dist} shows the average distances to the target position in the last $20$ steps of both SPMPC and baseline. Using the Euclidean-distance based cost function (Eq. \ref{deterministic_cost_function}), SPMPC converged to better performances during RL process. {As a comparison, SPMPC could not learn to improve its performance using the Mahalanobis-distance based cost function (Eq. \ref{probabilistic_cost_function}). One possible reason is the RL process needs to explore unknown states as similar as other RL methods. Using the Mahalanobis-distance based cost function, the agent may avoid to transit to the states with large uncertainties and therefore learned a local minimum with poor control performance since the GP model initialized by a limited number of samples that could not sufficiently explore enough states for achieving the task.}

    {Turning to the successful learning result using the Euclidean-distance based cost function in Fig.\ref{fig:cov_dist}, when $N_{initial}=1$, SPMPC iteratively improved its performance from $196.53$ $m$ to $72.20$ $m$ within 10 iterations to be close to the baseline's, $54.66$ $m$, since the GP model initialized by few samples results in a poor exploration of RL. With more initial random samples, SPMPC accelerated the learning process to outperform the baseline, the average distance was to the target was decreased from $66.74$ $m$ to $29.4$ $m$ ($N_{initial}=5$), $42.32$ $m$ to $27.79$ $m$ ($N_{initial}=10$).}
    Figure \ref{fig:pid} gives examples of learning result of SPMPC and the baseline. Both the manually tuned baseline and the proposed method performed well under small disturbances. With larger disturbances, only the proposed method could successfully drive the boat to the target. The PID controller could not reach and stay close to the target even with a higher control frequency than SPMPC due to its disability of handling disturbances.

    {The model prediction accuracy was evaluated by testing the model at each iteration on additional $40$ rollouts with random control signals and wind, current settings ($L_{rollout}=50$). With $50$ initial samples, SPMPC decreased the average prediction error (with $95\%$ confidence interval) from $9.87\pm15.55$ $m$ to $7.88\pm11.21$ $m$ within $10$ iterations. With $250$ initial samples, the average error decreased from $6.65\pm9.6$ $m$ to $6.35\pm8.58$ $m$. With $500$ initial samples, the average error decreased from $6.29\pm7.97$ $m$ to $6.26\pm8.01$ $m$. Although the prediction error reached a lower limit and stop decreasing with $500$ initial samples due to the strong disturbances and bias described in Section \ref{S4}, it is reasonable compared with the boat length in our simulation ($7.9$ $m$).}
    
    The first experiment empirically confirmed that using the Euclidean-distance based cost function, the proposed RL approach can converge over the iterations with a very limited number of samples, and achieve better control performance than the baseline. {It also indicated the importance of both initial samples and RL exploration in SPMPC. The number of initial samples affects the model prediction accuracy and the quality of RL exploration. The RL process explores to collect samples that focus on reducing the task-specific cost function. As an example shown in Fig. \ref{fig:RL_explore}, the $500$ initial samples gives the agent sufficient knowledge of driving behaviors. Then the RL process iteratively explores unknown states to finally achieve the task. }
    
    \begin{figure}
      \begin{subfigure}[b]{0.66\columnwidth}
        \includegraphics[width=1.0\linewidth]{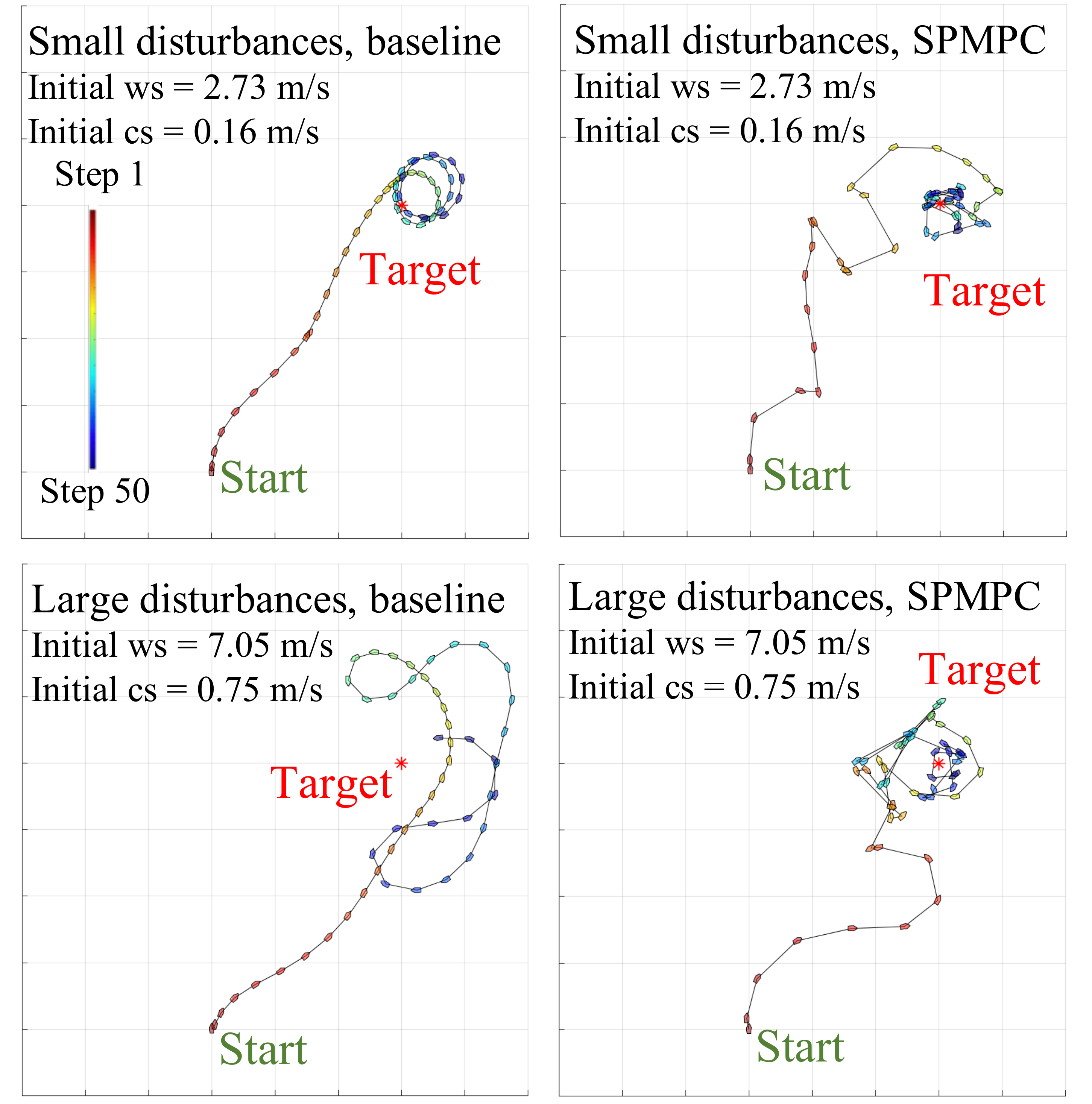}
        \caption{}
        \label{fig:pid_compare}
      \end{subfigure}%
      \begin{subfigure}[b]{0.325\columnwidth}
        \includegraphics[width=1.0\linewidth]{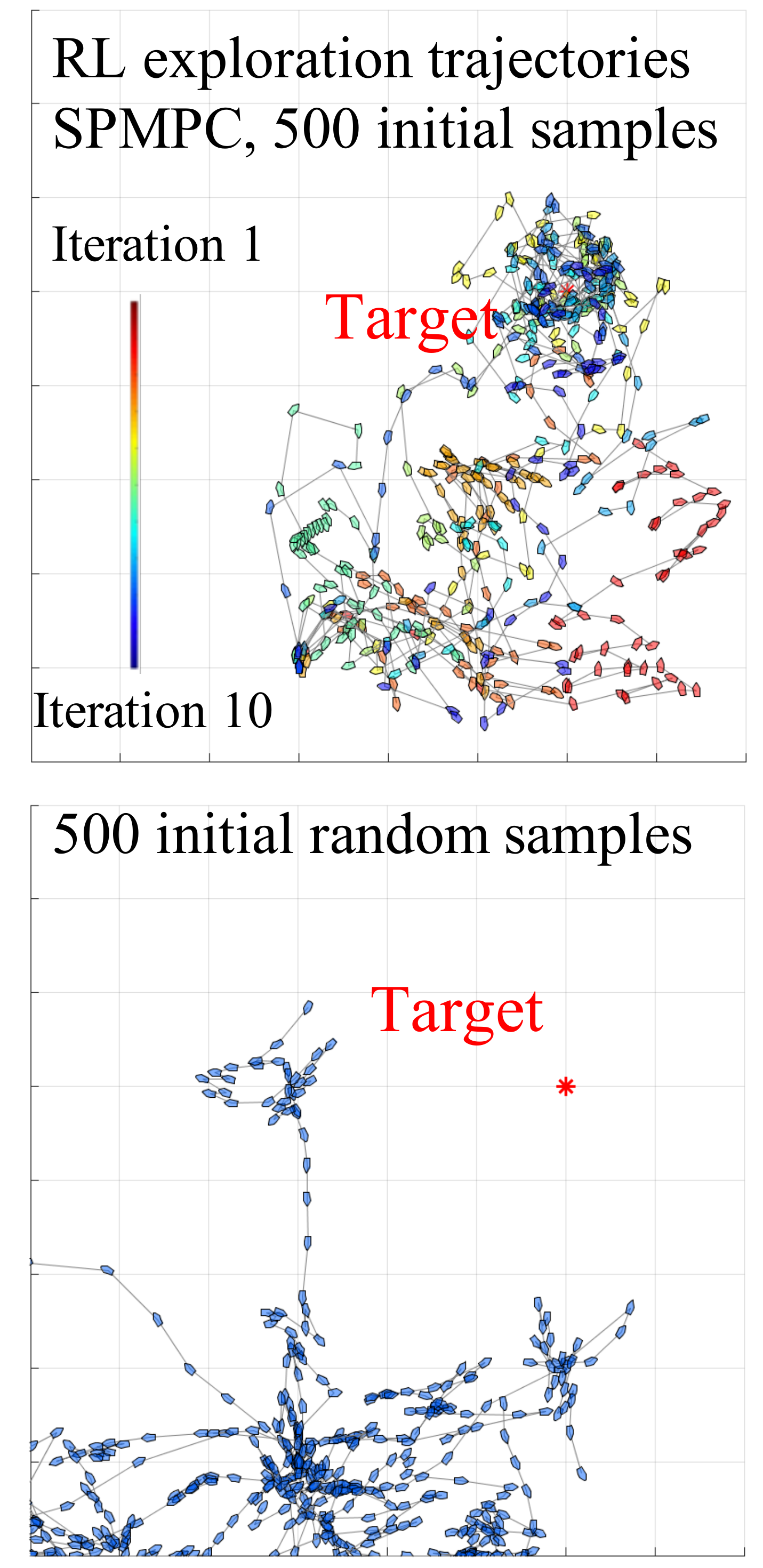}
        \caption{}
        \label{fig:RL_explore}
      \end{subfigure}%
      \caption{Examples of (a) the baseline and SPMPC in simulation task (b) the RL exploration samples and initial samples}
      \label{fig:pid}
    \end{figure}

    \begin{figure}
    \begin{center}
    \includegraphics[width=1\columnwidth]{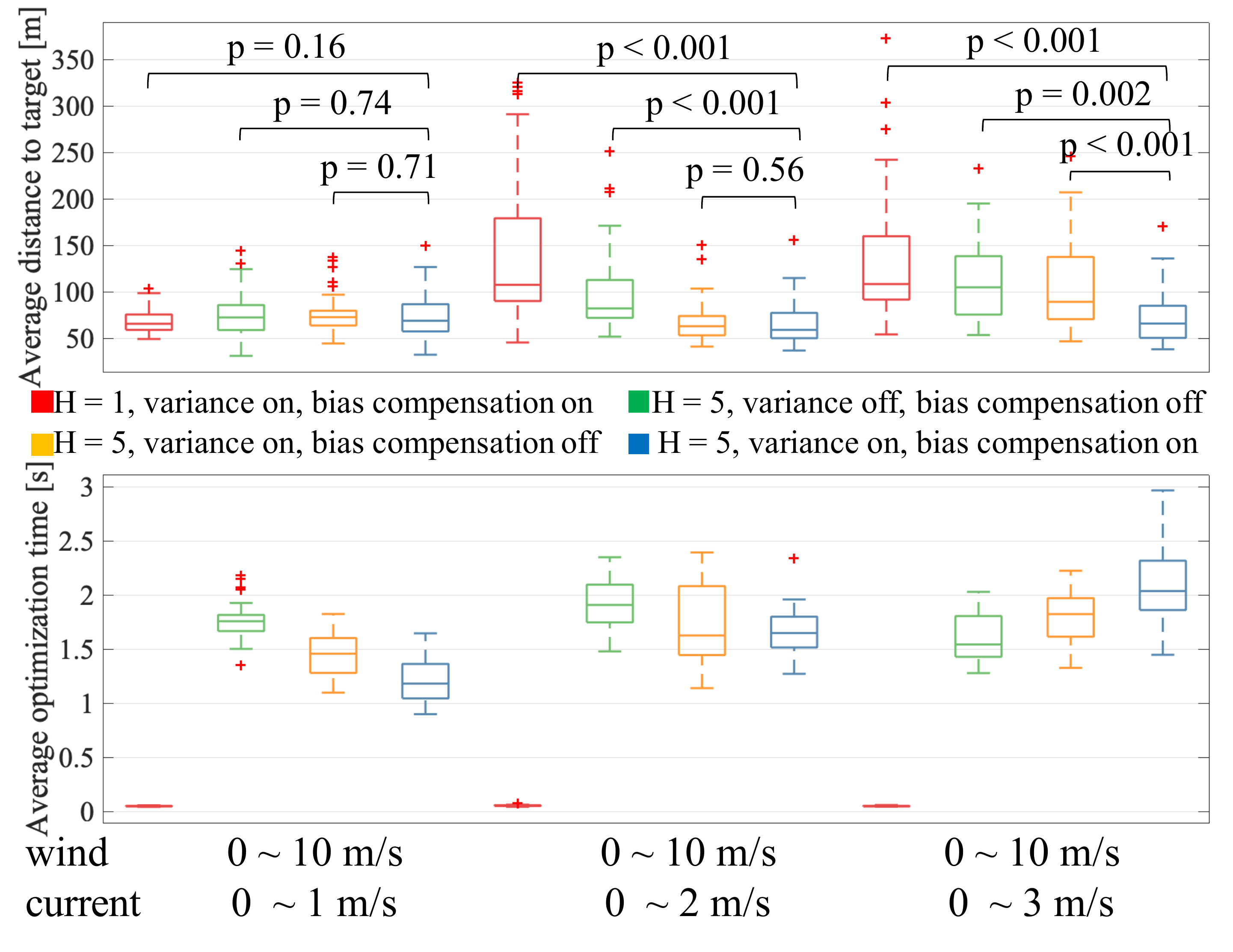}
    \caption{Control performance and optimization time over 50 tests with different settings.}
    \label{fig:simu_com}
    \end{center}
    \end{figure}

    \begin{figure}
    \begin{center}
    \includegraphics[width=0.8\columnwidth]{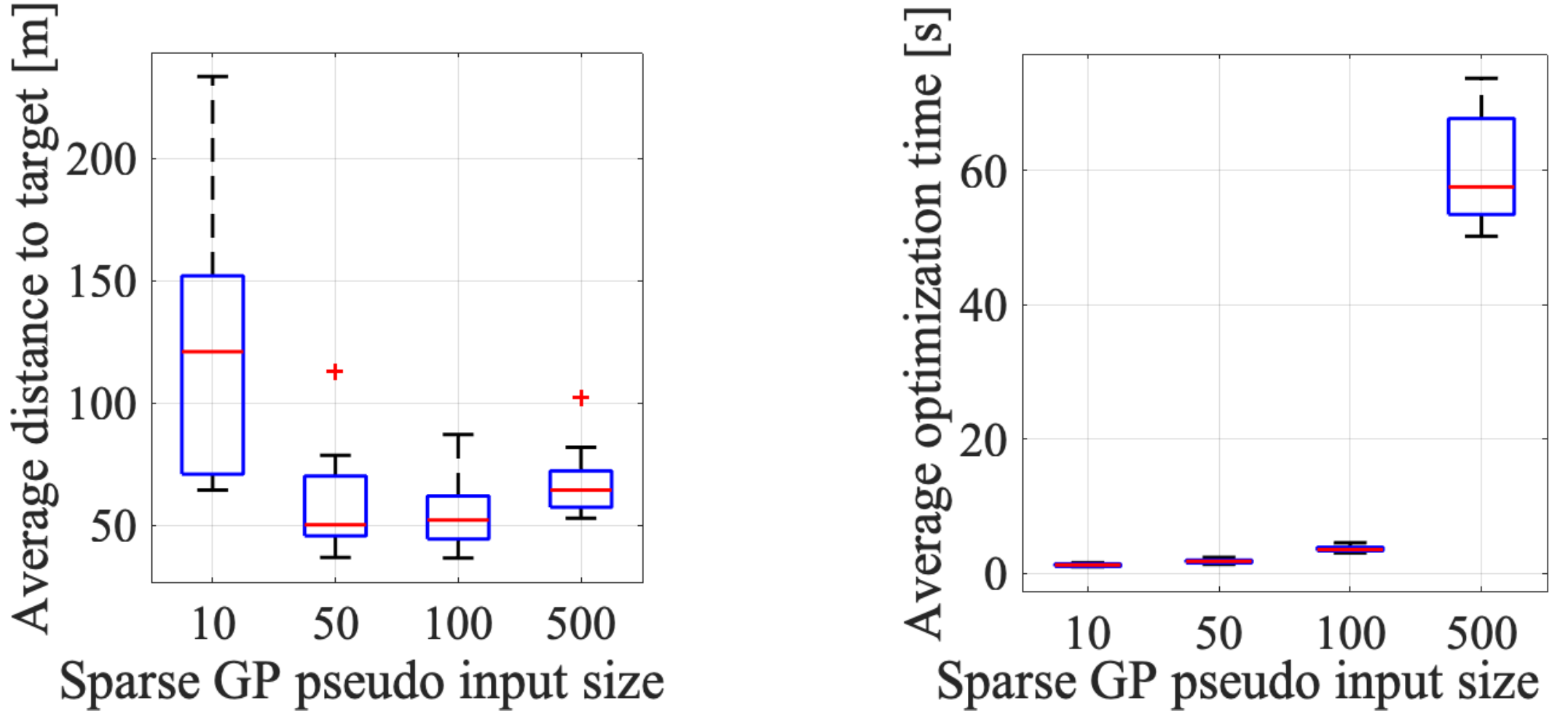}
    \caption{The effect of sparse GP setting on control performance and optimization time.}
    \label{fig:sp}
    \end{center}
    \end{figure}

    \subsubsection{The effect of SPMPC Settings}\label{S5-1-3}
    
    {
    The second experiment is to evaluate 1) whether the longer horizon contributes to better control results, 2) whether considering the uncertainties of predicted state, i.e. the boat position, velocity, and direction improve performance in a challenging environment, 3) whether the bias compensation contributes to better control performance. Four different configurations of SPMPC are compared:
    \begin{enumerate}
        \item $H = 1$, variance on, bias compensation on
        \item $H = 5$, variance off, bias compensation on
        \item $H = 5$, variance on, bias compensation off
        \item $H = 5$, variance on, bias compensation on
    \end{enumerate}
    Most settings follow Section \ref{S5-1-2} with Euclidean-distance based cost function. The max speeds of current are set to $1, 2, 3$ $m/s$ as a parameter of unobservable uncertainty. The experiment starts with $N_{initial}=10$ rollouts with random actions to train a GP model, followed by $N_{trials}=10$ rollouts to iteratively update the GP model. The learned controller is tested with another $50$ rollouts.
    }

{The average distances to the target position and optimization times over $L_{rollout}=50$ steps are shown in Fig. \ref{fig:simu_com}. Note that an average bias near $70$ $m$ is a good result since the start position is over $400$ $m$ far away to the target. All four settings worked well with a small current. On the other hand, configuration 4 outperformed others as the max current speed increased. The t-test result (with significance level $\alpha$ = 0.05) on different settings in Fig. \ref{fig:simu_com} indicates that the long prediction horizon, uncertainty information (variance) in long-term prediction and bias compensation contribute to a significantly better and more robust control performance in a challenging ocean environment.} On the other hand, searching each step control signal takes approx. $0.05$ $s$ while long horizon prediction takes over $1$ $s$. With the max current speed of $3$ $m/s$, configuration 4 takes the longest time (approx $2$ $s$) per search to perform well. These results indicate the ability of SPMPC system to drive the boat to the target position with a high degree of sample efficiency ($1000$ samples) and robustness in a challenging environment.
    
    
    \subsubsection{The effect of Sparse GP Pseudo-inputs}\label{S5-1-4}
    
    {The last experiment is to investigate the effect of the sparse GP, i.e. the trade-off between control performance and computational efficiency. We take configuration 4 (five step prediction, support variance, and bias compensation) with $10, 50, 100$ and $500$ sparse GP pseudo-inputs and set the max current speed to $2$ $m/s$, $N_{rollout}=10$ and $N_{trials}=10$. The learned GP model is tested with another $10$ rollouts.}

    {
    According to Fig. \ref{fig:sp}, $50$ pseudo-inputs performs slightly worse (average $121.03$$m$ v.s. $64.46$ $m$) than $500$ pseudo-inputs but is far faster (average $1.55$$s$ v.s. $57.57$$s$). This result indicate the potential of efficiently running SPMPC with sparse GP.
    }
    
    
    
    

    \begin{figure}
    \begin{center}
    \includegraphics[width=1\columnwidth]{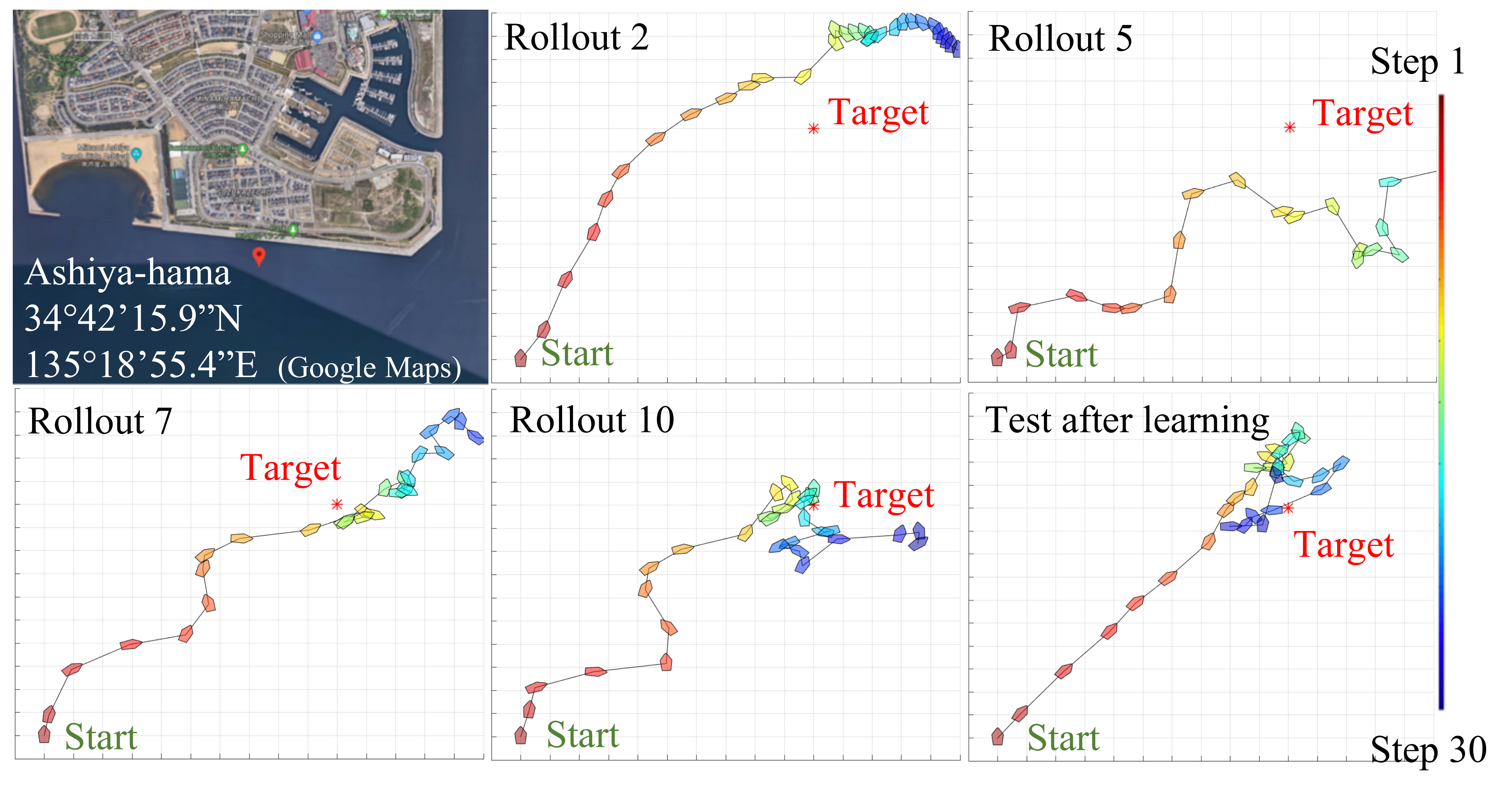}
    \caption{The area for real experiment, routes of the real boat during the RL process and an example trial after learning}
    \label{fig:real_demo_route_map}
    \end{center}
    \end{figure}

    \subsection{Real Experiments}\label{S5-2}
    In this section, we apply SPMPC system to a real boat autopilot task. All hardware (Sec. \ref{S4-1}) was provided by FURUNO ELECTRIC CO., LTD. {States and actions in this experiment are defined in Table \ref{tab:sim_state_action}.} {The experimental area is Ashiya-hama, Ashiya, Hyogo, Japan (\ang{34;42;15.9}N \ang{135;18;55.4}E, left top of Fig. \ref{fig:real_demo_route_map}) with the corresponding weather conditions: cloudy, current speed $0.0 \sim 0.2 m/s$, current direction \ang{45}, wave height $0.3 \sim 0.5 m$, wind speed $2 m/s$, wind direction \ang{135}. This weather information is averaged over a large area, while in reality both wind and current experienced by the boat continuously change as it traverses through.} 
    
    {For the autopilot task, one rollout is defined as $L_{rollout}=30$ steps. It started from the initial position $[0, 0]$ with initial orientation close to \ang{0}. The objective is to reach $[100, 100]$ and remain as close as possible.} At each step, the control signal was operated for around $7$ s, including $5$ s operation time and $2$ s optimization time $t_{opt}$ {(unlike in the simulation, the real optimization time here was strictly limited to $t_{opt}$)}. The GP model was initialized by $N_{initial}=10$ random rollouts. Then $N_{trial}=10$ rollouts were applied in the RL process. The SPMPC settings were $H=5$, variance on and bias compensation on, with the Euclidean-distance based cost function and Sparse GP (50 pseudo-inputs).} 
    

    During the RL process, the SPMPC system successfully learned to reach the target position within $20$ steps and attempted to remain near that position against disturbances from the real ocean environment. Several route maps during RL process are shown in Fig. \ref{fig:real_demo_route_map}. One example of SPMPC with $10$ rollouts learning is shown in the right bottom of Fig. \ref{fig:real_demo_route_map}\footnote{video is available at \url{https://youtu.be/jOpw2cFP0mo}}. With a total $600$ steps samples ($50\%$ from random sampling and $50\%$ from RL), the boat reached the target position and stay within $30$ steps (about $3.5$ minutes). These results indicate the proposed system is able to accomplish the real-sized boat autopilot task in a real ocean environment. Without any human demonstration, the RL process iteratively learned a robust MPC controller resilient to strong disturbances such as wind and current, with a high degree of sample efficiency and reasonable calculation times.
    
    \section{Discussion and Future Work}\label{S6}
    
    This work presents SPMPC, an RL approach specialized for autonomous boat which is challenging due to the strong and unpredictable disturbances in the ocean environment and the extremely high cost of getting learning samples with the real boat. SPMPC combines the model-based RL, GP model and MPC framework together to naturally handle the real-time ocean uncertainties with the efficiency of both calculation and sampling. A system based on SPMPC is successfully applied it to the autopilot task of a real-size boat in both simulation and real ocean environment with not only robustness to disturbances but also great sample efficiency and quick calculation with a limited computational resource.  

    

    {{For the future work, we will compare the proposed method with related work \cite{kamthe2018data} to further investigate its efficiency in simulation and the real ocean.} Since multiple GP models are trained for each target dimension in our present setting following \cite{deisenroth2013gaussian}, it may be beneficial to apply the multi-dimensional output GP \cite{alvarez2010efficient} that reduces the computational complexity of training, multiple models since the control frequency in the proposed system is limited due to the trade-off between optimization quality and computational complexity. Because the current system is developed in Matlab (autonomous system) and Labview (boat hardware interface), general computational bottlenecks could be improved by moving to C++ and CUDA for future developments. We also believe that the performance of our system would be improved by the addition of a suitable current sensor. Another topic of interest is to directly learn expert driving skills by building a GP model-based on human demonstrations, to potentially achieve a more human-like autonomous driving.  
    }

    
    \bibliographystyle{ieeetr}
    \bibliography{paper}

\begin{thebibliography}{10}

\bibitem{fagnant2015preparing}
D.~J. Fagnant and K.~Kockelman, ``Preparing a nation for autonomous vehicles:
  opportunities, barriers and policy recommendations,'' {\em Transportation
  Research Part A: Policy and Practice}, vol.~77, pp.~167--181, 2015.

\bibitem{pastore2010improving}
T.~Pastore and V.~Djapic, ``Improving autonomy and control of autonomous
  surface vehicles in port protection and mine countermeasure scenarios,'' {\em
  Journal of Field Robotics}, vol.~27, no.~6, pp.~903--914, 2010.

\bibitem{tomic2012toward}
T.~Tomic, K.~Schmid, P.~Lutz, A.~Domel, M.~Kassecker, E.~Mair, I.~L. Grixa,
  F.~Ruess, M.~Suppa, and D.~Burschka, ``Toward a fully autonomous {UAV}:
  Research platform for indoor and outdoor urban search and rescue,'' {\em IEEE
  robotics \& automation magazine}, vol.~19, no.~3, pp.~46--56, 2012.

\bibitem{sutton1998reinforcement}
R.~S. Sutton and A.~G. Barto, {\em Reinforcement learning: An introduction}.
\newblock MIT press Cambridge, 1998.

\bibitem{kober2013reinforcement}
J.~Kober, J.~A. Bagnell, and J.~Peters, ``Reinforcement learning in robotics: A
  survey,'' {\em The International Journal of Robotics Research}, vol.~32,
  no.~11, pp.~1238--1274, 2013.

\bibitem{7989202}
G.~Williams, N.~Wagener, B.~Goldfain, P.~Drews, J.~M. Rehg, B.~Boots, and E.~A.
  Theodorou, ``Information theoretic {MPC} for model-based reinforcement
  learning,'' in {\em International Conference on Robotics and Automation
  (ICRA)}, pp.~1714--1721, 2017.

\bibitem{tran2015reinforcement}
L.~D. Tran, C.~D. Cross, M.~A. Motter, J.~H. Neilan, G.~Qualls, P.~M. Rothhaar,
  A.~Trujillo, and B.~D. Allen, ``Reinforcement learning with autonomous small
  unmanned aerial vehicles in cluttered environments,'' in {\em Aviation
  Technology, Integration, and Operations Conference}, p.~2899, 2015.

\bibitem{liu2016unmanned}
Z.~Liu, Y.~Zhang, X.~Yu, and C.~Yuan, ``Unmanned surface vehicles: An overview
  of developments and challenges,'' {\em Annual Reviews in Control}, vol.~41,
  pp.~71--93, 2016.

\bibitem{naeem2012colregs}
W.~Naeem, G.~W. Irwin, and A.~Yang, ``Colregs-based collision avoidance
  strategies for unmanned surface vehicles,'' {\em Mechatronics}, vol.~22,
  no.~6, pp.~669--678, 2012.

\bibitem{lefeber2003tracking}
E.~Lefeber, K.~Y. Pettersen, and H.~Nijmeijer, ``Tracking control of an
  underactuated ship,'' {\em IEEE transactions on control systems technology},
  vol.~11, no.~1, pp.~52--61, 2003.

\bibitem{annamalai2015robust}
A.~S. Annamalai, R.~Sutton, C.~Yang, P.~Culverhouse, and S.~Sharma, ``Robust
  adaptive control of an uninhabited surface vehicle,'' {\em Journal of
  Intelligent \& Robotic Systems}, vol.~78, no.~2, pp.~319--338, 2015.

\bibitem{peng2013adaptive}
Z.~Peng, D.~Wang, Z.~Chen, X.~Hu, and W.~Lan, ``Adaptive dynamic surface
  control for formations of autonomous surface vehicles with uncertain
  dynamics,'' {\em IEEE Transactions on Control Systems Technology}, vol.~21,
  no.~2, pp.~513--520, 2013.

\bibitem{polydoros2017survey}
A.~S. Polydoros and L.~Nalpantidis, ``Survey of model-based reinforcement
  learning: Applications on robotics,'' {\em Journal of Intelligent \& Robotic
  Systems}, vol.~86, no.~2, pp.~153--173, 2017.

\bibitem{rasmussen2006gaussian}
C.~E. Rasmussen and C.~K. Williams, {\em Gaussian processes for machine
  learning}, vol.~1.
\newblock MIT press Cambridge, 2006.

\bibitem{ghavamzadeh2016bayesian}
M.~Ghavamzadeh, Y.~Engel, and M.~Valko, ``Bayesian policy gradient and
  actor-critic algorithms,'' {\em The Journal of Machine Learning Research},
  vol.~17, no.~1, pp.~2319--2371, 2016.

\bibitem{Martin2018SparseGP}
J.~Martin, J.~Wang, and B.~Englot, ``Sparse gaussian process temporal
  difference learning for marine robot navigation,'' in {\em CoRL}, 2018.

\bibitem{deisenroth2013gaussian}
M.~P. Deisenroth, D.~Fox, and C.~E. Rasmussen, ``Gaussian processes for
  data-efficient learning in robotics and control,'' {\em IEEE Transactions on
  Pattern Analysis \& Machine Intelligence}, vol.~37, no.~2, pp.~408--423,
  2013.

\bibitem{cao2017gaussian}
G.~Cao, E.~M.-K. Lai, and F.~Alam, ``Gaussian process model predictive control
  of an unmanned quadrotor,'' {\em Journal of Intelligent \& Robotic Systems},
  vol.~88, no.~1, pp.~147--162, 2017.

\bibitem{kamthe2018data}
S.~Kamthe and M.~Deisenroth, ``Data-efficient reinforcement learning with
  probabilistic model predictive control,'' in {\em International Conference on
  Artificial Intelligence and Statistics}, pp.~1701--1710, 2018.

\bibitem{mackay2003information}
D.~J. MacKay, {\em Information theory, inference and learning algorithms}.
\newblock Cambridge university press, 2003.

\bibitem{girard2003gaussian}
A.~Girard, C.~E. Rasmussen, J.~Q. Candela, and R.~Murray-Smith, ``Gaussian
  process priors with uncertain inputs application to multiple-step ahead time
  series forecasting,'' in {\em Advances in neural information processing
  systems}, pp.~545--552, 2003.

\bibitem{deisenroth2009analytic}
M.~P. Deisenroth, M.~F. Huber, and U.~D. Hanebeck, ``Analytic moment-based
  {G}aussian process filtering,'' in {\em the annual international conference
  on machine learning}, pp.~225--232, 2009.

\bibitem{nocedal2006sequential}
J.~Nocedal and S.~J. Wright, ``Sequential quadratic programming,'' {\em
  Numerical optimization}, pp.~529--562, 2006.

\bibitem{mayne2000constrained}
D.~Q. Mayne, J.~B. Rawlings, C.~V. Rao, and P.~O. Scokaert, ``Constrained model
  predictive control: Stability and optimality,'' {\em Automatica}, vol.~36,
  no.~6, pp.~789--814, 2000.

\bibitem{snelson2006sparse}
E.~Snelson and Z.~Ghahramani, ``Sparse {G}aussian processes using
  pseudo-inputs,'' in {\em Advances in neural information processing systems},
  pp.~1257--1264, 2006.

\bibitem{alvarez2010efficient}
M.~{\'A}lvarez, D.~Luengo, M.~Titsias, and N.~Lawrence, ``Efficient multioutput
  gaussian processes through variational inducing kernels,'' in {\em
  International Conference on Artificial Intelligence and Statistics},
  pp.~25--32, 2010.

\end{thebibliography}
    
    \end{document}